# Superconducting phases in potassium intercalated iron selenides

Tianping Ying, Xiaolong Chen*, Gang Wang*, Shifeng Jin, Xiaofang Lai, Tingting Zhou, Han Zhang, Shijie Shen, Wanyan Wang

Research & Development Center for Functional Crystals, Beijing National Laboratory for Condensed Matter Physics, Institute of Physics, Chinese Academy of Sciences, Beijing 100190, China

*Supporting Information Placeholder*

**ABSTRACT:** The ubiquitous coexistence of majority insulating 245 phases and minority superconducting phases in $A_xFe_{2-y}Se_2$ (A = K, Cs, Rb, Tl/Rb, Tl/K) by high-temperature routes makes pure superconducting phases highly desirable for studying the intrinsic properties of this superconducting family. Here we report that there at least exist two pure superconducting phases $K_xFe_2Se_2(NH_3)_y$ ($x \approx 0.3$, and $0.6$) determined only by potassium concentration in the potassium-intercalated iron selenides using the liquid ammonia route. $K_{0.3}Fe_2Se_2(NH_3)_{0.47}$ corresponds to the 44 K phase with $c = 15.56(1)$ Å and $K_{0.6}Fe_2Se_2(NH_3)_{0.37}$ to the 30 K one with $c = 14.84(1)$ Å. By further increasing potassium doping, the 44 K phase can be converted into the 30 K phase. The ammonia has little, if any, effect on superconductivity. Therefore, the conclusion should apply to $K_{0.3}Fe_2Se_2$ and $K_{0.6}Fe_2Se_2$ superconducting phases. $K_{0.3}Fe_2Se_2(NH_3)_{0.47}$ and $K_{0.6}Fe_2Se_2(NH_3)_{0.37}$ stand out among known superconductors as their structures are stable only at particular potassium doping levels, and hence the variation of $T_c$ with doping is not dome-like.

Considerable progresses have been made in the study of $A_xFe_{2-y}Se_2$ (A = K, Cs, Rb, Tl/Rb, Tl/K) since the discovery of superconductivity at about 30 K in $K_{0.8}Fe_2Se_2$.[1] The dominant phases in $A_xFe_{2-y}Se_2$ are now identified to be $A_2Fe_4Se_5$ (so called 245 phase) that exhibit disorder-order structural transitions due to Fe vacancies followed by antiferromagnetic orderings with a large magnetic moment up to 3.3 $\mu_B$ per iron ion at $T_N$ =478 K, 534 K and 559 K for A = Cs, Rb, and K, respectively.[2] These 245 phases, however, are insulating and not responsible for the observed superconductivity. Studies by transmission electron microscopy,[3] synchrotron X-ray diffraction,[4] Mössbauer spectroscopy,[5] nuclear magnetic resonance,[6] muon spin spectroscopy,[7] scanning tunneling microscopy,[8] infrared spectroscopy[9] and Raman spectroscopy[10] give strong evidences that phase separations occur in $A_xFe_{2-y}Se_2$, leading to 245 phases with $\sqrt{5} \times \sqrt{5} \times 1$ superstructure and superconducting phases coexist. Alterations of initial compositions, heating procedures, and cooling rates are proved not to prevent the phase separations.[11]

The superconducting phases are generally thought to precipitate from $A_xFe_{2-y}Se_2$ into nanoscale strips that intergrow with the 245 phases[4] though other formation routes cannot be ruled out. Their volume fractions are agreeably low, 10-20% estimated by various techniques.[4-7] All experimental results suggest that they be isostructural to $BaFe_2As_2$[12] with Fe vacancy free and can be stood for by a chemical formula $A_xFe_2Se_2$.[3c,4,6,8] The deduced alkali occupancies, however, are inconsistent, varying a lot from x = 0.3 to 1.0[4,6,8] because of lacking single phase samples. Efforts to obtain pure or superconducting-dominated phases by using Bridgeman or other high-temperature routes have failed up to now. In addition, a 44 K superconducting phase with trace fractions was often observed apart from the 30 K phase in some samples,[11a,13] further complicating the identification of superconducting phases. Thus, one has good reasons to think the previously reported properties of this family of superconductors are flawed or even dubious considering the fact that the 'single crystal' samples, on which measurements were done, actually consist of majority 245 phases and minority superconducting phases. Therefore, it is a pressing and important issue to obtain pure superconducting phases or real single crystals for clarifying their peculiar electronic structures[14] and studying the intrinsic properties of this new family of superconductors.

**Table 1. K, Fe, and Se nominal compositions and sample compositions characterized by ICP-AES and standard Micro-Kjeldahl method.**

| Sample | Nominal composition | Composition by ICP-AES and Micro-Kjeldahl method |
|---|---|---|
| 1 | $K_{0.2}Fe_2Se_2+NH_3$ | $K_{0.2}Fe_{2.03}Se_2(NH_3)_{0.45}$ |
| 2 | $K_{0.25}Fe_2Se_2+NH_3$ | $K_{0.24}Fe_{1.98}Se_2(NH_3)_{0.49}$ |
| 3 | $K_{0.27}Fe_2Se_2+NH_3$ | $K_{0.26}Fe_{2.01}Se_2(NH_3)_{0.45}$ |
| 4 | $K_{0.3}Fe_2Se_2+NH_3$ | $K_{0.28}Fe_{1.97}Se_2(NH_3)_{0.47}$ |
| 5 | $K_{0.35}Fe_2Se_2+NH_3$ | $K_{0.33}Fe_{1.96}Se_2(NH_3)_{0.4}$ |
| 6 | $K_{0.4}Fe_2Se_2+NH_3$ | $K_{0.4}Fe_2Se_2(NH_3)_{0.5}$ |
| 7 | $K_{0.45}Fe_2Se_2+NH_3$ | $K_{0.44}Fe_{1.98}Se_2(NH_3)_{0.4}$ |
| 8 | $K_{0.5}Fe_2Se_2+NH_3$ | $K_{0.48}Fe_{2.02}Se_2(NH_3)_{0.35}$ |
| 9 | $K_{0.6}Fe_2Se_2+NH_3$ | $K_{0.6}Fe_2Se_2(NH_3)_{0.37}$ |
| 10 | $K_{0.65}Fe_2Se_2+NH_3$ | $K_{0.62}Fe_2Se_2(NH_3)_{0.35}$ |
| 11 | $K_{0.7}Fe_2Se_2+NH_3$ | $K_{0.57}Fe_{2.02}Se_2(NH_3)_{0.34}$ |
| 12 | $K_{0.8}Fe_2Se_2+NH_3$ | $K_{0.68}Fe_2Se_2(NH_3)_{0.3}$ |

Previously, we reported the synthesis of a series of superconductors with $T_c$ = 30~46 K by intercalating metals Li, Na, K, Ba, Sr, Ca, Eu and Yb in between FeSe layers via an ammonothermal route.[15] This route enabled us to tune the concentrations of intercalated metals at room temperature, implying an alternative for preventing the phase separations. In this Communication, we report that in the potassium-intercalated iron selenides, there at least exist two superconducting phases, with difference mainly in potassium concentrations. Both phases are Fe vacancy free and can be represented by $K_xFe_2Se_2(NH_3)_y$ with $x \approx 0.3$ corresponding to the 44 K phase and $x \approx 0.6$ to the 30 K one. We show the inserted ammonia has little, if any, effect on superconductivity. No antiferromagnetic orderings are observed for both phases. Unlike iron pnictide superconductors, only two discrete potassium doping levels are allowed to stabilize the structure and the trend of $T_c$ variation with doping levels does not exhibit a dome-like curve in the selenide superconductors.

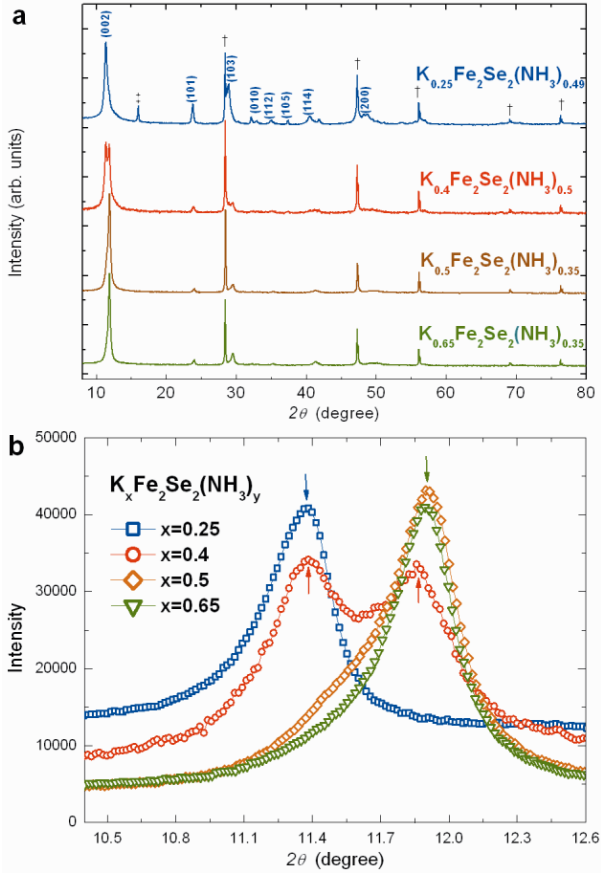

**Fig 1.** (a) Powder X-ray diffraction patterns for samples with nominal composition $K_xFe_2Se_2(NH_3)_y$ (x = 0.25, 0.4, 0.5, and 0.65, respectively) measured at 297 K, Cu $K\alpha$ radiation. The patterns were referenced against Si powder (NIST SRM 640, denoted with '†'). For x = 0.25, residual $\beta$-FeSe noted with '‡' is observed. With increasing potassium content, the lattice with larger cell constant $c$ weakens and another lattice with smaller $c$ develops. The coexistence of these two phases is clearly seen at x = 0.4. For x = 0.4, 0.5, and 0.65, no peaks corresponding to $\beta$-FeSe are observed. (b) The enlargement of (002) peaks. The peaks are marked with arrows. Unless otherwise specified, x in this figure and hereafter are nominal compositions for clarity.

A series of samples $K_xFe_2Se_2(NH_3)_y$ with nominal x = 0.1, 0.2, 0.25, 0.27, 0.3, 0.35, 0.4, 0.45, 0.5, 0.6, 0.7, 0.8 and 1.0 were synthesized by a liquid ammonia method. The synthesizing details are described in Supporting Information. The key to obtain phase pure sample is to use high pure FeSe precursors. The existence of impurities will capture the solvated electrons first and impede potassium intercalating. The chemical analyses were done by inductively coupled plasma-atomic emission spectrometry (ICP-AES) for potassium, Fe and Se. Cautions should be taken to avoid the escape of $H_2Se$ during measurements. This often leads to underestimating Se contents. Chemical analyses by ICP-AES indicate Fe:Se is very close to 1:1 for all thus-obtained samples within an error less than 2%, as shown in Table 1, strongly indicating these samples are Fe vacancy free. We found the potassium amounts in samples match well with predetermined initial amounts within variations less than 7% when x < 0.6. Potassium amounts in samples remain at about 0.6 despite further increasing initial potassium amounts. We determined the ammonia amounts in these compounds by standard Micro-Kjeldahl method (see Table 1). The ammonia amount, y, varies in samples and all are less than 0.5, roughly in an inverse proportion with potassium contents. Phase identifications were performed on an X-ray diffractometer with Cu $K\alpha$ radiation. The indexing was aided by Si standards to obtain precise lattice constants. The magnetic properties were characterized using a physical property measurement system (Quantum Design) and a vibrating sample magnetometer.

Fig. 1(a) shows the powder X-ray diffraction patterns for nominal $K_xFe_2Se_2(NH_3)_y$ samples with x = 0.25, 0.4, 0.5, and 0.65. Peaks marked with '†' are from Si (Standard Reference Material, SRM640), which was added in all samples to correct the diffraction peak positions. All patterns can be indexed based on a tetragonal body-centered cell with lattice constants $a$ = 3.82~3.85 Å and $c$ = 14.8~15.7 Å, indicating varying amounts of potassium are intercalated in between the FeSe layers. We note that (00ℓ) diffractions are most well-defined among all peaks due to preferred orientations. So, they are good indicators of phase identifications and variations of the lattice constants $c$ with intercalated potassium. From the magnified pattern for (002) in Fig.1 (b), we can clearly see peak shifts with the potassium intercalations. A striking feature is that the peak shifts are not continuous with the potassium contents. Instead, peaks only appear at about $2\theta$ = 11.36 ° and 11.89 °, corresponding to $c$ = 15.56(1) Å and 14.84(1) Å in all samples, respectively. For instance, $K_{0.25}Fe_2Se_2(NH_3)_{0.49}$ exhibits one (002) peak at $2\theta$ = 11.36 °, and $K_{0.65}Fe_2Se_2(NH_3)_{0.35}$ at $2\theta$ = 11.89 °, indicating these samples consist only of one potassium-intercalated phases with different constants $c$. In contrast, $K_{0.4}Fe_2Se_2(NH_3)_{0.5}$ has two (002) peaks, meaning it contains both the two phases.

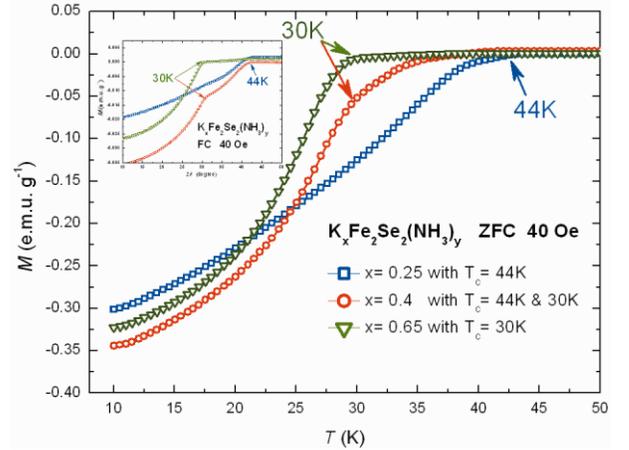

**Fig. 2.** ZFC M-T curves of samples with different nominal potassium contents. When x = 0.25 and 0.4, onset transition temperature ($T_c$) at 44 K is observed. For x = 0.65, $T_c$ shifts to 30 K. The inset shows the transition region for the corresponding FC M-T curves. An inflexion point at 30 K can be clearly seen for $K_{0.4}Fe_2Se_2(NH_3)_{0.5}$.

The magnetism of all samples were measured over a temperature range from room temperature to 10 K. Shown in Fig. 2 are typical M-T curves under zero field cooling (ZFC) for $K_{0.25}Fe_2Se_2(NH_3)_{0.49}$, $K_{0.4}Fe_2Se_2(NH_3)_{0.5}$, and $K_{0.65}Fe_2Se_2(NH_3)_{0.35}$. The onset transition temperatures are 44 K, 44 K and 30 K, respectively. An inflexion point at 30 K can be clearly seen in the field cooling (FC) curve for $K_{0.4}Fe_2Se_2(NH_3)_{0.5}$ (the inset), implying two superconducting phases coexist. While no apparent inflexions are observed in $K_{0.25}Fe_2Se_2(NH_3)_{0.49}$ and

$K_{0.65}Fe_2Se_2(NH_3)_{0.35}$, they can be regarded 44 K and 30 K phase dominated, respectively. In addition, inflexion points at 36 K often appear in some samples with x ≈ 0.35~0.55. All data for the transition temperatures $T_c$ and the lattice constants $c$ are summarized in Fig. 3(a) and 3(b). With x < 0.3, all samples exhibit onset $T_c$ at 44 K and lattice constants $c$ of 15.56(1) Å, while with x > 0.6, all samples exhibit onset $T_c$ at 30 K and $c$ of 14.84(1) Å. In between x ≈ 0.3 and x ≈ 0.6, the samples' superconducting transition start at 44 K, but with distinct inflexion points at 36 K or 30 K, indicating the samples contain two or three superconducting phases. The lattice constant $c$ relating to the 36 K phase, however, is not observed. The reason is unknown at the present. To verify $T_c$ is potassium dependent, we increase the potassium doping amount by further reacting $K_{0.3}Fe_2Se_2(NH_3)_{0.47}$ with potassium in liquid ammonia. Chemical analyses confirm that the potassium concentration indeed increases from 0.28 to 0.57 while almost no compositional change in Fe and Se after reacting. A dramatic change in superconductivity ensues. Fig. 4 shows the M-T curve for the $K_{0.3}Fe_2Se_2(NH_3)_{0.47}$ sample before and after further reacting. It can be clearly seen that the 44 K phase is converted into the 30 K phase.

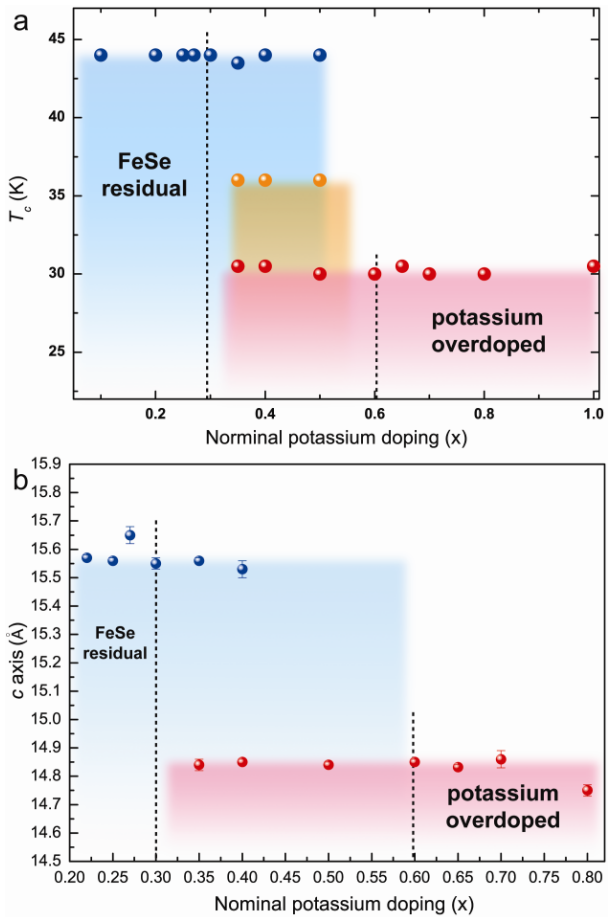

**Fig. 3.** (a) Superconducting transition temperatures of $K_xFe_2Se_2(NH_3)_y$ as a function of nominal potassium contents. $T_c$ show abrupt change against the potassium contents rather than dome-like. With x < 0.3, there is residual β-FeSe which could be distinguished from XRD patterns. With x > 0.6, the unique blue color induced by potassium dissolved in liquid ammonia does not fade within several minutes and potassamide is finally obtained, sticking on the tube wall. The transition at 36 K does not exist alone and is always accompanied with transitions at 44 K or 30 K. (b) Cell constants $c$ of $K_xFe_2Se_2(NH_3)_y$ vs nominal potassium contents. Two distinct $c$ are displayed. For x located between 0.3 and 0.6, these two $c$ coexist.

These results demonstrate that there at least exist two superconducting phases in the potassium-intercalated iron selenides, one being close to $K_{0.3}Fe_2Se_2(NH_3)_{0.47}$ with $T_c$ = 44 K and another close to $K_{0.6}Fe_2Se_2(NH_3)_{0.37}$ with $T_c$ = 30 K. These two phases are isostructural to each other, except for the former phase's lattice $a$ shrinking a little and $c$ expanding in comparison with the latter's. The enhanced $T_c$ from 30 k to 44 K is attributed to the decrease in intercalated potassium amount from 0.6 to 0.3, implying the 30 K phase is overdoped by electrons. Most importantly, this study reveals a striking feature of $K_xFe_2Se_2(NH_3)_y$ superconductors that the variation of $T_c$ vs potassium doping is not continuous but discrete, simply because the stable phases only appear at several doping levels. This feature is unique to $K_xFe_2Se_2(NH_3)_y$ as all cuprate and pnictide superconductors show a dome-like variations of $T_c$ with doping and so do the variations of doped phases' lattice constants.

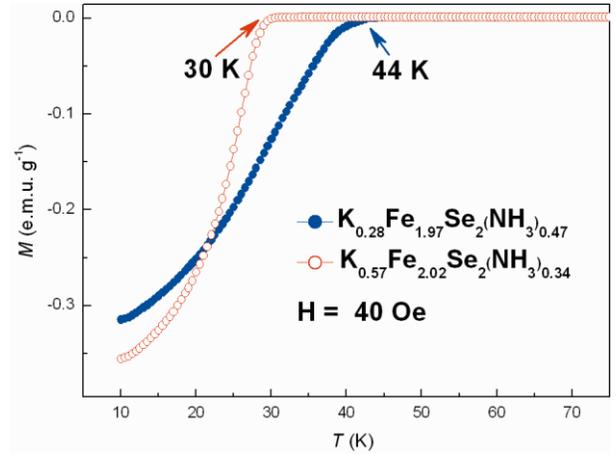

**Fig. 4.** ZFC M-T curves of $K_{0.28}Fe_{1.97}Se_2(NH_3)_{0.47}$ before and after further reacting. For $K_{0.28}Fe_{1.97}Se_2(NH_3)_{0.47}$, transition at 44 K is observed. After further reacting with potassium in liquid ammonia, $K_{0.28}Fe_{1.97}Se_2(NH_3)_{0.47}$ is converted into $K_{0.57}Fe_{2.02}Se_2(NH_3)_{0.34}$. The transition temperature is down to 30 K. The chemical formulas in this figure are real compositions.

The lattice constants $c$ = 15.56(1) Å for $K_{0.3}Fe_2Se_2(NH_3)_{0.47}$ and $c$ = 14.81(1) Å for $K_{0.6}Fe_2Se_2(NH_3)_{0.37}$ are much longer than c ≈ 14.2 Å[3c,4,8] for the superconducting phase without ammonia, suggesting in between FeSe layers there exist other atoms or atomic groups as we previously pointed out in Ref. [15]. By using similar method, Burrard-Lucas and coworkers[16] intercalated Li into FeSe layers and found lithium amide and ammonia exist apart from lithium ions in product compound characterized by neutron diffractions. Scheidt and coworkers[17] ruled out the existence of lithium amide in between FeSe layers based on their own experiments. To study the effect of $NH_3$ on structure and superconductivity, we performed the extraction of ammonia while avoiding the collapse of the lattice. To this end, we found the optimal extraction temperature is about 200 °C. The as-prepared $K_{0.3}Fe_2Se_2(NH_3)_{0.47}$ sealed in a tube was first placed in an oil bath while slowly heating to 200 °C then evacuated for several hours. Powder X-ray diffraction pattern (see Fig.S1 in supporting information) shows that the tetragonal body-centered phase remains while lattice $a$ expanding from 3.843(1) Å to 3.87(1) Å and lattice $c$ shrinking a lot from 15.56(1) Å to 14.28(4) Å after heating process. The $c$ value is very close to the reported one for the superconducting phase obtained by high-temperature routes.[3c,4,8] The temperature-dependant magnetization in Fig. S2 shows the onset transition temperature is still close to 44 K. Similar situation is found in $K_{0.6}Fe_2Se_2(NH_3)_{0.37}$. These results demonstrate the ammonia has little, if any, influence on $T_c$. Therefore, although our results are obtained based on the $NH_3$ containing samples, the conclusion should apply to $K_{0.3}Fe_2Se_2$ and $K_{0.6}Fe_2Se_2$ supercon-ducting phases though ammonia causes changes of lattice constants.

Recently, Yan and Miao[18] pointed out, based on the first-principles calculations, the 40 K phase and 30 K phase in KFe$_2$Se$_2$ are due to different occupying sites of potassium in the $I4/mmm$ lattice, leading to $c$ = 18.6-19.04 Å and $c$ = 15.46-16.7 Å, respectively. Our results do not support their calculations as we think that the intercalated potassium amounts that occupy the body-centered site play a crucial role in determining the superconducting phases. The elongated lattice constants $c$ are attributed to inserted ammonia or other atomic groups.[15-17,19]

Last, we point out all these superconducting phases are paramagnetic above $T_c$, agreeing with the recently reported results.[7] Show in Fig. S3 is the magnetization vs temperature from 10 K to 400 K under a field of 1 T for K$_{0.27}$Fe$_2$Se$_2$(NH$_3$)$_{0.45}$. The magnetization value and its trend vs temperature reveal that the sample contain no Fe vacancy phases such as 245 phase or Fe$_7$Se$_8$, agreeing well with our results of chemical analyses that the obtained superconducting phases are Fe vacancy free.

In summary, in the potassium-intercalated iron selenides, there at least exist two superconducting phases with ThCr$_2$Si$_2$ structure type. Both FeSe layers are vacancy free, but with different potassium concentrations. It is the potassium concentration that determines the superconducting transition temperatures. K$_{0.3}$Fe$_2$Se$_2$(NH$_3$)$_{0.47}$ and K$_{0.6}$Fe$_2$Se$_2$(NH$_3$)$_{0.37}$ stand out among known superconductors in that their structures are stable only at certain potassium doping levels, and hence the variation of $T_c$ with doping is not dome-like either. The mechanism behind these unique properties needs further investigation to understand.

## ASSOCIATED CONTENT

### Supporting Information

Supporting Information accompanies this paper is available at http://pubs.acs.org.

## AUTHOR INFORMATION

### Corresponding Author

X.L.C. (email: chenx29@iphy.ac.cn) or G.W. (email: gangwang@iphy.ac.cn).

### Author Contributions

These authors contributed equally.

### Notes

The authors declare no competing financial interest.

## ACKNOWLEDGMENT

T. P. Ying thanks P. Zheng and Y. G. Shi from Institute of Physics, CAS for help with part of magnetic measurements. This work was partly supported by the National Natural Science Foundation of China under Grant Nos. 90922037, 51072226 and 51202286, by the Chinese Academy of Sciences, and ICDD.